\documentclass[preprint, superscriptaddress,preprintnumbers,amsmath,amssymb,pra]{revtex4}
\usepackage{graphicx}

\usepackage{graphicx}

\usepackage{color}

\begin{document}

\newcommand{\be}{\begin{equation}}
\newcommand{\ee}{\end{equation}}

\thispagestyle{empty}

\title{How to observe the giant thermal effect in the Casimir
force for graphene systems}

\author{
G.~Bimonte}
\affiliation{Dipartimento di Fisica E. Pancini, Universit\`{a} di Napoli
Federico II, Complesso Universitario MSA, Via Cintia, I-80126 Napoli, Italy}
\affiliation{INFN Sezione di Napoli, I-80126, Napoli, Italy}

\author{
G.~L.~Klimchitskaya}
\affiliation{Central Astronomical Observatory at Pulkovo of the
Russian Academy of Sciences, Saint Petersburg,
196140, Russia}
\affiliation{Institute of Physics, Nanotechnology and
Telecommunications, Peter the Great Saint Petersburg
Polytechnic University, Saint Petersburg, 195251, Russia}

\author{
V.~M.~Mostepanenko}
\affiliation{Central Astronomical Observatory at Pulkovo of the
Russian Academy of Sciences, Saint Petersburg,
196140, Russia}
\affiliation{Institute of Physics, Nanotechnology and
Telecommunications, Peter the Great Saint Petersburg
Polytechnic University, Saint Petersburg, 195251, Russia}
\affiliation{Kazan Federal University, Kazan, 420008, Russia}

\begin{abstract}
A differential measurement scheme is proposed which allows
for a clear observation of the giant thermal effect for the Casimir
force, that was recently predicted to occur in graphene systems
at short separation distances. The difference among the Casimir
forces acting between a metal-coated sphere and the two halves of
a dielectric plate, one uncoated and the other coated with graphene,
is calculated in the framework of the Dirac model using
the rigorous formalism of the polarization tensor. It is
shown that in the proposed configuration both the difference
among the Casimir forces and its thermal contribution
can be easily measured using already existing experimental
setups. An observation of the giant thermal effect should
open opportunities for modulation and control of dispersion
forces in micromechanical systems based on
graphene and other novel 2D-materials.
\end{abstract}

\maketitle

\section{INTRODUCTION}

During the last few years graphene attracted a widespread
attention in physics, material science and nanotechnological
applications due to its unusual mechanical, optical and
electrical properties \cite{1,2,3,4}. These properties
originate from the fact that graphene is a two-dimensional
layer of carbon atoms organized in a honeycomb crystal
lattice. At low energies the quasiparticles in graphene are
massless (or almost massless) charged fermions described by
the Dirac equation. Interest in graphene and relativistic
Dirac-like systems in condensed matter physics further
increased after other  two-dimensional honeycomb materials
called silicene, germanene and stanene (which are 2D allotropes
of Si, Ge and Sn) have been discovered \cite{5,6,7}.

Another much studied interdisciplinary subject  is
represented by the
fluctuation-induced (dispersion) forces acting between two
microparticles, a material surface and a microparticle, and
between two closely spaced material surfaces. These forces
are of entirely quantum origin. For separations between
the interacting bodies smaller than a few nanometers these
interactions have a
nonrelativistic character and are commonly known as
{\it van der Waals forces } \cite{8}. At larger separations
relativistic effects come into play and the name
{\it Casimir forces} \cite{9} is most often used.
Even though the energies associated with these forces usually
represent just a small contribution to the total
free energy of the respective material systems, they are
presently considered to be a crucial ingredient for a
qualitatively correct and quantitatively accurate
description of the binding properties of materials \cite{10}.
This is the reason why  van der Waals and Casimir forces
are actively studied both experimentally and theoretically
(see Refs.~\cite{11,12,13} for a review). In so doing,
puzzling findings concerning the role of dissipation of free
charge carriers in metals
have been reported \cite{14,15,16,17,18,19}
and important applications of dispersion forces in
micromechanical systems \cite{20,21,22} and nanochips
\cite{23} have been proposed.

In view of the universal character of dispersion
forces, it is widely acknowledged
that their role is also important
for closely spaced graphene sheets, graphene-coated
substrates, and graphene and regular 3D-materials.
The van der Waals and Casimir forces in microsystems involving
graphene have already been studied using a variety of
theoretical approaches \cite{24,25,26,27,28,29,30,31,32,33,34}.
In the framework of the Dirac model \cite{1,2,3,4}, the
natural description of the Casimir force in graphene
systems, based on the first principles of quantum electrodynamics
at nonzero temperature, is provided by the formalism of the
polarization tensor in (2+1)-dimensional space-time
\cite{35,36,37,38,39,40,41}. In the one-loop
approximation this approach is indeed equivalent \cite{42}
to the formalism
of density-density correlation functions in the
random-phase approximation. Using the latter approach, it was
discovered \cite{24} that differently from 3D-materials
the Casimir force for graphene systems displays a giant
thermal effect, which provides  a very large  contribution to the
force magnitude at relatively small separations of about
150\,nm. This is surprising if one considers that for
3D-materials the thermal effect becomes dominant only for
separations larger than the thermal length
$\lambda_T=\hbar c/(k_BT)$ (about $7\,\mu$m for room
temperature), where the magnitude of the Casimir force
is anyhow very small.
Detailed investigations of the thermal effect in graphene
systems \cite{43,44} demonstrated that it can be harnessed
to modulate and control  the strength of the
Casimir force. In spite of its
importance and unusual features, this
thermal effect has not been
measured yet.

In this paper, we propose an experimental approach allowing
for an unambiguous measurement of the giant thermal effect in the
Casimir force between a Au-coated sphere and a graphene-coated
Si substrate made of dielectric Si. According to our results,
this can be achieved by measuring the {\it difference}
among the Casimir
forces between a Au-coated sphere and the two halves of a Si
plate, one of which is coated and the other is uncoated with
a graphene sheet (see Fig.~1). A conceptually similar differential approach
has been
successfully used in the past to determine the role of
relaxation of free charge carriers in the Casimir force
between metallic test bodies \cite{19,45,46,47} and for
constraining hypothetical corrections of Yukawa type
to Newton's law of gravitation \cite{48,49}. The
experimental setups used in these investigations
achieved a force sensitivity of a fraction
of 1\,fN at room temperature. Below we show that at separations of 100\,nm,
1 and
$1.5\,\mu$m our measurement scheme leads to differential
Casimir forces equal to approximately 2.3\,pN, 18.5\,fN,
and 8.1\,fN,
respectively. At the same separations, according to our
results, the thermal effect in graphene contributes
1.5\,pN, 17.5\,fN, and 7.8\,fN, respectively.
Thus, this effect should be easily
measurable over the entire separation range from 100\,nm to
$1.5\,\mu$m taking into account the sensitivity of existing
experimental setups.

The structure of the paper is as follows. In Sec. II we describe the proposed experimental setup and
show how to compute the Casimir force difference between the two halves of the Si plate
 on the basis of the
formalism of the polarization tensor for graphene. In Sec. III we report the results of our numerical computations.
Section~IV  presents our conclusions and an outline for future work.

\section{ EXPERIMENTAL SCHEME AND GENERAL FORMALISM }

We consider the configuration of a Au-coated sphere of
radius $R=150\,\mu$m moving back and forth in vacuum at some
height $a$ above a plate made of dielectric Si (see Fig.~1). The thickness
of the Au
coating is chosen to be large enough  (typically larger than
a few tens of nanometers) that the
sphere can be considered  as made entirely of Au for
the sake of computing the
Casimir force.
One half of the plate (the right half in Fig.~1) is coated with a graphene sheet.
In a concrete experimental implementation, one could use
a setup similar to those described in
Refs.~\cite{19,48,49} employing a micromechanical torsional
oscillator. Setups of this sort provide a direct
measurement of the difference $F_{\rm diff}(a,T)$ among the Casimir forces
$F_{\rm Si}(a,T)$ and $F_{\rm gr}(a,T)$ acting, respectively, between a sphere and the uncoated and graphene-coated halves
of a Si plate:
\be
F_{\rm diff}(a,T)=F_{\rm Si}(a,T)-F_{\rm gr}(a,T)\;.\label{fdiff}
\ee

The force difference $F_{\rm diff}$  at $T=300\,$K
can be calculated using the Lifshitz formula \cite{9,51}.
We assume that in Eq. (\ref{fdiff})  $F_{\rm Si}(a,T)$ and $F_{\rm gr}(a,T)$ are the forces acting  on  the sphere when its tip is placed above points   that are respectively deep in the left and right halves of the Si plate,  in such a way that the effect of the sharp boundary between the graphene-coated half and the uncoated half of the Si plate can be neglected.  In practice, to achieve this,  it is sufficient  to place  the sphere tip at a distance $\delta$ from the boundary that is larger than a few times the typical Casimir interaction radius $\rho=\sqrt{a R}$. Under these conditions the forces $F_{\rm gr}(a,T)$ and $F_{\rm Si}(a,T)$ become respectively undistinguishable from the forces that would act between the Au sphere and two distinct  {\it homogeneous} Si plates, one  fully covered with graphene  and the other fully uncovered. The latter forces can be computed using
the proximity force approximation \cite{52}
(in Refs.~\cite{52a,53,54,55} it was rigorously proved that
 this approximation is
sufficiently precise under the condition $a\ll R$). The result is
\begin{equation}
F_{\rm diff}(a,T)=\frac{k_BTR}{4a^2}
\sum_{l=0}^{\infty}{\vphantom{\sum}}^{\prime}
\int_{\zeta_l}^{\infty}\!\!ydy
\sum_{\alpha}
\ln\frac{1-r_{\alpha}^{(1)}(i\zeta_l,y)r_{\alpha}^{(2)}(i\zeta_l,y)
e^{-y}}{1-r_{\alpha}^{(1)}(i\zeta_l,y)R_{\alpha}(i\zeta_l,y)
e^{-y}},
\label{eq1} 
\end{equation}
\noindent
where $k_B$ is the Boltzmann constant,
the prime in the first summation sign means that the term with $l=0$
is taken with weight 1/2, $\zeta_l=2a\xi_l/c$
 are the dimensionless Matsubara frequencies,
the dimensional Matsubara frequencies being defined as
$\xi_l=2\pi k_BTl/\hbar$, and
$y=2a\sqrt{\mbox{\boldmath{$k$}}_{\bot}^2+{\xi_l^2}/{c^2}}$
 with
the in-plane wave vector $\mbox{\boldmath{$k$}}_\bot$.
The summation in $\alpha$ in Eq.~(\ref{eq1}) is over the
 transverse magnetic ($\alpha=\mbox{TM}$) and
transverse electric ($\alpha=\mbox{TE}$) polarizations of the
electromagnetic field. The reflection coefficients
$r_{\alpha}^{(n)}$ on the
boundaries between vacuum and Au $(n=1)$ or Si $(n=2)$ are
given by
\begin{eqnarray}
&&
r_{\rm TM}^{(n)}(i\zeta_l,y)=\frac{\varepsilon_l^{(n)}y-
k_l^{(n)}}{\varepsilon_l^{(k)}y+k_l^{(n)}},
\nonumber \\
&&
r_{\rm TE}^{(n)}(i\zeta_l,y)=
\frac{y-k_l^{(n)}}{y+k_l^{(n)}},
\label{eq2}
\end{eqnarray}
\noindent
where $\varepsilon_l^{(n)}\equiv\varepsilon^{(n)}(i\zeta_l)$
are the dielectric permittivities of Au and Si calculated
at the imaginary Matsubara frequencies and
$k_l^{(n)}=[y^2+(\varepsilon_l^{(n)}-1)\zeta_l^2]^{1/2}$.
Finally, the reflection coefficients $R_{\alpha}$ on the boundaries
between vacuum and the graphene-coated substrate (Si) can be
expressed via the polarization tensor of graphene using
the results of Refs.~\cite{37,42,56,57}. The same results were
obtained in Ref.~\cite{40} in the framework of another
approach. In terms of the dimensionless variables one
has \cite{43}
\begin{equation}
R_{\rm TM}(i\zeta_l,y)=\frac{\varepsilon_l^{(2)}y-
k_l^{(2)}+yk_l^{(2)}(y^2-\zeta_l^2)^{-1}
\tilde{\Pi}_{00,l}}{\varepsilon_l^{(2)}y+k_l^{(2)}+
yk_l^{(2)}(y^2-\zeta_l^2)^{-1}\tilde{\Pi}_{00,l}},
\qquad
R_{\rm TE}(i\zeta_l,y)=\frac{y-k_l^{(2)}-(y^2-\zeta_l^2)^{-1}
\tilde{\Pi}_l}{y+k_l^{(2)}-(y^2-\zeta_l^2)^{-1}
\tilde{\Pi}_l}.
\label{eq3}
\end{equation}
\noindent
Here, the dimensionless polarization tensor of graphene
calculated at the Matsubara frequencies, $\tilde{\Pi}_{mn,l}$,
is expressed via the dimensional one as
$\tilde{\Pi}_{mn,l}\equiv\tilde{\Pi}_{mn}(\zeta_l,y)=
2a\Pi_{mn,l}/\hbar$ and the quantity $\tilde{\Pi}_l$ is
defined by
\begin{equation}
\tilde{\Pi}_l\equiv(y^2-\zeta_l^2)\tilde{\Pi}_{{\rm tr},l}-
y^2\tilde{\Pi}_{00,l},
\label{eq4}
\end{equation}
\noindent
where $\tilde{\Pi}_{\rm tr}$ is the trace of the
polarization tensor. Note that the polarization tensor
describes the response of a physical system to an
electromagnetic field. Because of this, it is immediately
connected with physical quantities such as the dielectric
permittivity \cite{26,42} and conductivity \cite{58,59}
of graphene. For instance, the in-plane component of the
nonlocal dielectric permittivity of graphene calculated
at the imaginary Matsubara frequencies is given by \cite{26,42}
\begin{equation}
\varepsilon_l^{\|}=1+\frac{1}{2k_{\bot}\hbar}\Pi_{00,l}.
\label{eq5}
\end{equation}

The dielectric permittivities of Au and dielectric Si
entering Eqs.~(\ref{eq2}) and (\ref{eq3}) are found from
the optical data for the complex indices of refraction
(see Refs.~\cite{60} and \cite{61}, respectively) using
the Kramers-Kronig relation.
Since optical data for the
complex index of refraction of all materials
are available only for sufficiently large frequencies, at low frequencies the available data have to be supplemented with
some theoretical model.  When the relaxation properties of electrons
in metals are taken into account, the dielectric permittivity $\varepsilon_{l}^{(1)}$ of our material 1 (Au) at the Matsubara
frequencies can be represented in the form
\begin{equation}
\varepsilon_{l}^{(1,D)}=
\frac{\tilde{\omega}_{p,1}^2}{\zeta_l(\zeta_l+\tilde{\gamma_1})}
+\varepsilon_{1,l}^{\rm cor}.
\label{eq7bis}
\end{equation}
\noindent
Here, $\tilde{\omega}_{p,1}$ and $\tilde{\gamma}_1$ are,
respectively, the dimensionless plasma
frequency and relaxation parameter of Au connected with the
dimensional ones by
\begin{equation}
\tilde{\omega}_{p,1}=\frac{2a\omega_{p,1}}{c},\quad
\tilde{\gamma}_{1}=\frac{2a\gamma_{1}}{c}
\label{eq8bis}
\end{equation}
\noindent
and $\varepsilon_{1,l}^{\rm cor}$ is a contribution of the core (bound)
electrons to the dielectric permittivity determined by the optical data.
The upper index $D$ is used to stress that the permittivity (\ref{eq7bis})
has the Drude form. Note that the relaxation parameter $\tilde{\gamma}_1$
depends on temperature and goes to zero with vanishing $T$ by a power law
for metals with perfect crystal lattices. For real metals containing
some fraction of impurities there is rather small but nonzero residual
relaxation at $T=0$.
For Au the values of the plasma frequency
$\omega_{p,1}\approx 9\,\mbox{eV}=1.37\times 10^{16}\,$rad/s
and the relaxation parameter
$\gamma_{1}\approx 35\,\mbox{meV}=5.3\times 10^{13}\,$rad/s
have been used.
If relaxation properties of free electrons are neglected, the
dielectric permittivity of metals takes the plasma form
\begin{equation}
\varepsilon_{1,l}^{p}=
\frac{\tilde{\omega}_{p,1}^2}{\zeta_l^2}
+\varepsilon_{1,l}^{\rm cor}.
\label{eqplasma}
\end{equation}
\noindent
The plasma model is usually used in the region of infrared optics where
$\gamma_1\ll\xi_l$. For the high-resistivity Si considered in this paper the extrapolation to low frequencies  was done on the basis of the model
\begin{equation}
\varepsilon_{l}^{(2)}=
\varepsilon_{2,l}^{\rm cor}.
\label{eq12bis}
\end{equation}
\noindent
Note, that for Si
$\varepsilon_{2,0}^{\rm cor}\approx 11.67$.
As mentioned above, there
are puzzling results in the literature
\cite{14,15,16,17,18,19} concerning the role of dissipation
of free electrons in metals. According to these results,
measurements of Refs. \cite{14,15,16,17,18,19} are in
agreement with theoretical predictions of the Lifshitz
theory if the available optical data of Au are
extrapolated to zero frequency by means of the lossless
plasma model Eq.~(\ref{eqplasma}) and exclude predictions of the same theory
if the optical data are extrapolated by means of the
theoretically better motivated Drude model Eq.~(\ref{eq7bis}), which takes
the dissipation of free electrons into account.
In our configuration, however, the Drude-plasma dilemma
makes only a negligibly small influence on the
computational results \cite{37}. We will  explicitly see this in Sec.~III,
where we present our numerical results.

Another quantity entering Eq.~(\ref{eq3}) is the
polarization tensor of graphene. At $l=0$ the exact
expressions for the polarization tensor are given by
\cite{41,44}
\begin{eqnarray}
&&
\tilde{\Pi}_{00,0}=\frac{\pi\alpha y}{\tilde{v}_F}+
\frac{32\alpha}{\tilde{v}_F^2}
\frac{ak_BT}{\hbar c}\left(\ln 2-By\!\!
\int_0^1\!\frac{\sqrt{1-u^2}du}{e^{Byu}+1}\right),
\nonumber \\
&&
\tilde{\Pi}_{0}=\pi\alpha\tilde{v}_Fy^3-8\alpha
\tilde{v}_Fy^3\int_0^1\frac{u^2}{\sqrt{1-u^2}}
\frac{du}{e^{Byu}+1},
\label{eq6}
\end{eqnarray}
\noindent
where $\alpha$ is the fine-structure constant,
$\tilde{v}_F=v_F/c\approx 1/300$ is the reduced Fermi
velocity for graphene, and
$B\equiv \hbar c\tilde{v}_F/(4ak_BT)$.

For the sake of brevity, at $l\geq 1$ we present the
approximate expressions for the polarization tensor at
room temperature, which lead, however, to practically
exact results for the Casimir free energy at separations of
our interest exceeding 50\,nm \cite{44}
\begin{eqnarray}
&&
\tilde{\Pi}_{00,l}=
\frac{\alpha(y^2-\zeta_l^2)}{\sqrt{\tilde{v}_F^2y^2+\zeta_l^2}}
(\pi+Y_l),
\label{eq7} \\
&&
\tilde{\Pi}_{l}=
{\alpha(y^2-\zeta_l^2)}{\sqrt{\tilde{v}_F^2y^2+\zeta_l^2}}
(\pi+Y_l),
\nonumber
\end{eqnarray}
\noindent
where
\begin{equation}
Y_l=4\int_0^{\infty}\frac{du}{e^{\pi lu}+1}
\frac{u^2}{1+u^2}.
\label{eq8}
\end{equation}
\noindent
Note that the quantity $Y_l$ captures the explicit
dependence of the polarization tensor
on the temperature for $l\geq 1$, whereas an implicit
dependence on $T$
originates from the Matsubara frequencies.

We note also that the polarization tensor takes an exact account for
the nonzero relaxation properties of graphene carriers. This is seen
from the fact that along the real frequency axis this tensor does have
an imaginary part \cite{58,59} (see also Ref.~\cite{62} for another
theoretical approach to the description of relaxation of charge
carriers in graphene).

\section{NUMERICAL RESULTS}

We have performed numerical computations of the difference
$F_{\rm diff}$ among the Casimir forces using
Eqs.~(\ref{eq1})--(\ref{eq4}), (\ref{eq6}) and (\ref{eq7}).
The computational results are presented in Fig.~2(a)--(d)
as functions of separation. The top lines are computed
at $T=300\,$K and the bottom lines at $T=0\,$K
[in the latter case summation over the discrete
Matsubara frequencies in Eq.~(\ref{eq1}) is replaced
by an integration along the imaginary frequency axis;
the polarization tensor at $T=0\,$K is obtained from
Eq.~(\ref{eq7}) by omitting the quantity $Y_l$].
As is seen in Fig.~2, over the entire separation range the
difference  $F_{\rm diff}$ among the thermal Casimir forces
is large enough and can be measured by using already
existing experimental setups \cite{19,49}.

As we pointed out in Sec.~II, the force difference
$F_{\rm diff}$  is practically independent of the prescription used for the Au sphere, whether Drude or plasma. For example, for  the separations $a_1=100\,$nm and $a_2=800\,$nm
at $T=300\,$K
the plasma prescription gives $F_{\rm diff}(a_1)=2344.21\,$fN and  $F_{\rm diff}(a_2)=29.0373\,$fN, while with the Drude prescription we find respectively
$F_{\rm diff}(a_1)=2345.15\,$fN and  $F_{\rm diff}(a_2)=29.0282\,$fN. The reason for almost coinciding results provided by the two prescriptions is easily understood. Indeed the choice among the Drude and plasma prescriptions affects the Casimir force mainly via the TE $l=0$ mode.  Consider first the Drude prescription. With this prescription, the reflection coefficient of Au at zero frequency for TE polarization is zero, and then it follows that within the Drude prescription the $l=0$ TE mode contributes nothing to $F_{\rm diff}$, independently of the reflection coefficient of the Si plate.

Consider now the plasma prescription. With this other prescription, the $l=0$ TE reflection coefficient of Au is different from zero:
\begin{equation}
r_{{\rm TE},p}^{(1)}(0,y)=
\frac{y-\sqrt{y^2+\tilde{\omega}_{p,1}^2}}{y+
\sqrt{y^2+\tilde{\omega}_{p,1}^2}}\;.
\label{eq19bis}
\end{equation}
 For the uncoated half of the Si plate, according to Eq. (\ref{eq2}),
 the $l=0$ TE reflection coefficient is zero. This implies at once that the sphere-plate force $F_{\rm Si}$ receives no contribution from the $l=0$ TE mode, whatever prescription is used for Au.  Consider now the force $F_{\rm gr}$ acting on the sphere when its tip is above the graphene-coated half of the Si plate.    It turns out  that the zero-frequency component $\tilde{\Pi}_{0}$  of the polarization tensor, given in the second line of Eq.(\ref{eq6}),  is very small and according to the second of Eq. (\ref{eq3}) this in turn  implies that the $l=0$ TE reflection coefficient of the graphene-coated half of the Si plate is nearly zero. Because of that, the $l=0$ TE contribution to the force $F_{\rm gr}$, and then to  $F_{\rm diff}$,  is  negligibly small.  Summarizing  the above considerations, we find that  when the Drude prescription is used the contribution of the  $l=0$ TE mode to $F_{\rm diff}$ is zero, while with the plasma prescription it   is non-zero but negligible.  This explains why the force difference $F_{\rm diff}$ is practically insensitive to the prescription used for Au.

In Fig.~2 we display also the force difference $F_{\rm diff}(a,0)$ for $T=0$ which is computed using the zero-temperature expression for the graphene polarization tensor and the
 plasma model for Au.
 As it can be  seen from Fig.~2, the thermal correction $\Delta_T F_{\rm diff}$
\be
\Delta_T F_{\rm diff}=F_{\rm diff}(a,T)-F_{\rm diff}(a,0)\, ,
\ee which is equal to the difference between
the top and bottom lines in Fig.~2, contributes to $F_{\rm diff}$
significantly and can be easily determined from the
comparison between the measurement data and theory.
As an example, at separation distances $a=0.15$, 0.2, 0.5, 1.0,
and $1.5\,\mu$m the magnitudes of the thermal contributions
$\Delta_TF_{\rm diff}$ in $F_{\rm diff}$
are equal to 800, 406, 68.5, 17.5, and 7.8\,fN,
respectively, which are far larger than the current experimental
sensitivity.

It is interesting to determine the relative weights of the different
contributions which make up the thermal effect.
For this purpose,
we decompose the thermal correction $\Delta_T F_{\rm diff}$
to the differential force  into two parts
\begin{equation}
\Delta_{T}F_{\rm diff}(a,T)=\Delta_{T}^{\!(1)}F_{\rm diff}(a,T)
+\Delta_{T}^{\!(2)}F_{\rm diff}(a,T).
\label{eq26b}
\end{equation}
\noindent
Here, we have introduced the notations
\begin{eqnarray}
&&
\Delta_{T}^{\!(1)}F_{\rm diff}(a,T)=F_{\rm diff}(a,T)
-\bar{F}_{\rm diff}(a,T),
\nonumber \\
&&
\Delta_{T}^{\!(2)}F_{\rm diff}(a,T)=\bar{F}_{\rm diff}(a,T)
-F_{\rm diff}(a,0),
\label{eq26c}
\end{eqnarray}
\noindent
where $\bar{F}_{\rm diff}(a,T)$ is calculated by Eq.~(\ref{eq1})
 at $T=300\,$K, but  using the zero-temperature value for the graphene polarization tensor (keeping fixed the room temperature permittivities of Au and Si ).
 From Eq.~(\ref{eq26c}) it is clear
that $\Delta_{T}^{\!(1)}F_{\rm diff}$ originates from the explicit dependence of the  graphene polarization tensor on the temperature as a parameter, whereas $\Delta_{T}^{\!(2)}F_{\rm diff}$  represents
the contribution to the thermal
correction caused by the summation over the discrete Matsubara
frequencies and (if the Drude model is used)
by the temperature dependence of the  relaxation frequency  of Au.
In Fig.~3 we plot the ratio
$\Delta_TF_{\rm diff}/F_{\rm diff}$
 as a function
of separation (the solid line). In the same figure, the
short-dashed and long-dashed lines show, respectively, plots of the ratios $\Delta_{T}^{\!(1)}F_{\rm diff}/F_{\rm diff}$ and $\Delta_{T}^{\!(2)}F_{\rm diff}/F_{\rm diff}$   (the sum of
the short-dashed and long-dashed lines results in the
solid one). As it can be seen in Fig.~3, the total thermal
correction contributes to $F_{\rm diff}$ for fractions of
0.728, 0.781, 0.898, 0.946, and
0.963
at $a=0.15$, 0.2, 0.5, 1.0, and $1.5\,\mu$m, respectively.
The explicit dependence of the polarization tensor on $T$
by itself contributes for fractions of
0.506, 0.540, 0.613, 0.644, and 0.655
 at the same respective
separations. As to the thermal effect originating
 from the discrete
summation over the Matsubara frequencies using the
zero-temperature polarization tensor, it contributes
smaller fractions of $F_{\rm diff}$ equal to 0.222,
0.241, 0.285, 0.302, and 0.308 at separations
0.15, 0.2, 0.5, 1.0, and $1.5\,\mu$m, respectively.
If Au at low frequencies is described by the Drude model, this
leads to only negligibly small changes in the results obtained
which do not influence on the lines shown in Fig.~3.

\section{CONCLUDING REMARKS}

To conclude, we have proposed a Casimir configuration allowing
for an unambiguous measurement of the giant thermal effect
in graphene based on already existing experimental setups.
This conclusion is drawn in the framework of the Dirac
model of graphene using the rigorous formalism of the
polarization tensor based on the first principles of
quantum electrodynamics. According to our results,
both the differential force arising from the presence
of a graphene
coating on one half of a Si plate and the thermal
contribution to it are easily observable over the
considered separation region from 0.1 to $1.5\,\mu$m.
We have also shown that both the explicit dependence of
the material properties of graphene on the temperature,
as well as the
 implicit temperature dependence due to summation
over the Matsubara frequencies need to be taken into
account when comparing experiment with theory. The obtained
results open novel opportunities for the investigation
of dispersion interactions in graphene systems and can
be used for modulation and control of operational forces
in micromechanical devices which include
elements based on graphene and other 2D-materials.

\section*{Acknowledgments}
The work of V.M.M. was partially supported by the Russian
Government
Program of Competitive Growth of Kazan Federal University.

\newpage
\begin{figure}[b]
\vspace*{-8cm}
\centerline{\hspace*{-.5cm}
\includegraphics{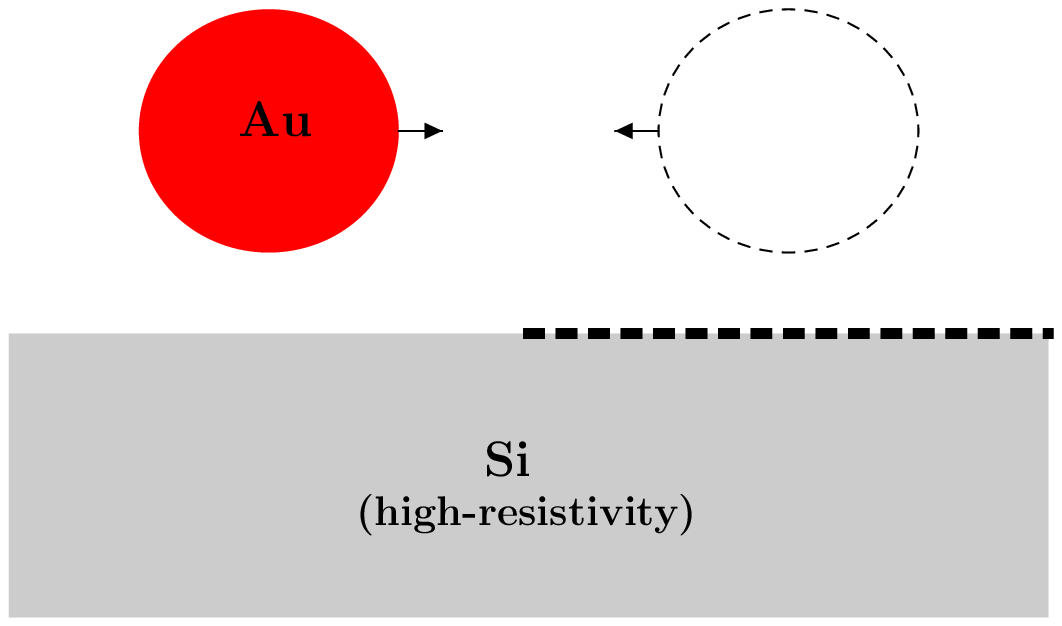}
}
\vspace*{-11cm}
\caption{\label{fg1}
The experimental configuration of a Au sphere moving back and
forth above a Si plate, half of which is covered with graphene (shown by dashes).
The measured quantity is the differential Casimir force $F_{\rm diff}$
between the Au sphere and the two halves of the plate when the sphere tip
is far away from their boundaries. The figure displays the two extreme
positions of the sphere during its motion.
The size of the sphere is shown not to scale.
}
\end{figure}
\begin{figure}[b]
\vspace*{-6cm}
\centerline{\hspace*{-.5cm}
\includegraphics{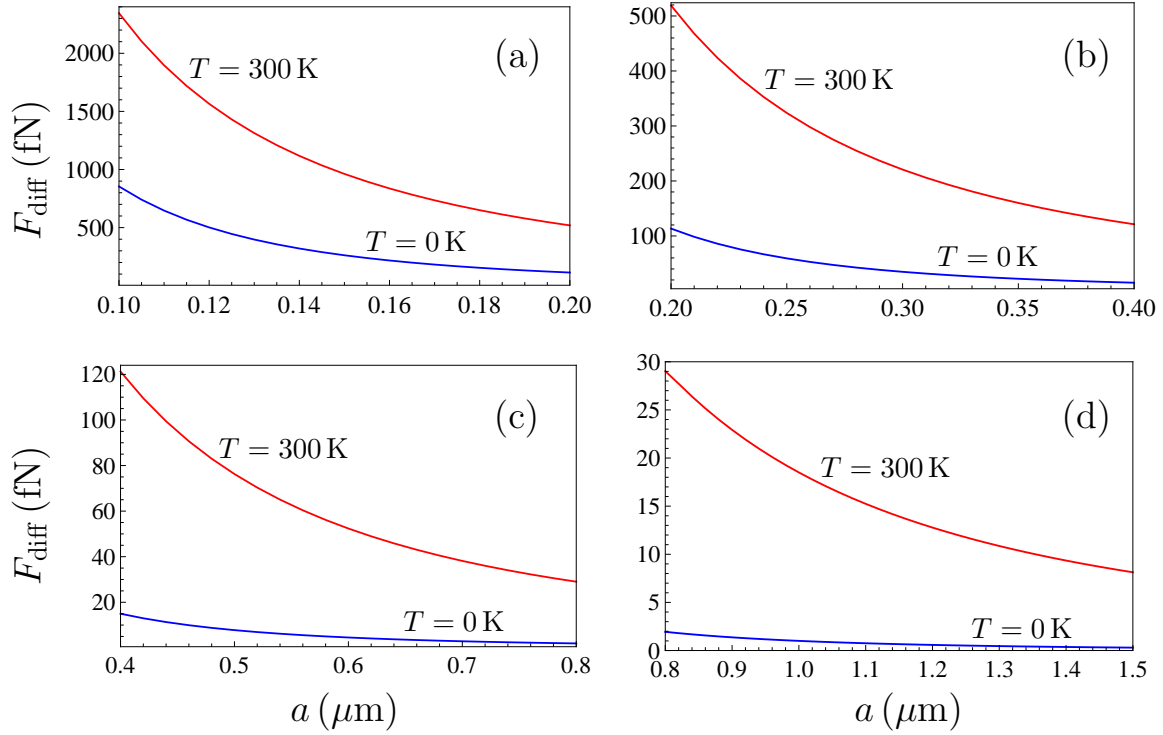}
}
\vspace*{-11cm}
\caption{\label{fg2}
The differences $F_{\rm diff}$ among the Casimir forces calculated at
$T=300\,$K (top lines) and at $T=0\,$K
(bottom lines) are shown as functions of
separation over four different separation intervals.
}
\end{figure}
\begin{figure}[b]
\vspace*{-8cm}
\centerline{\hspace*{3.5cm}
\includegraphics{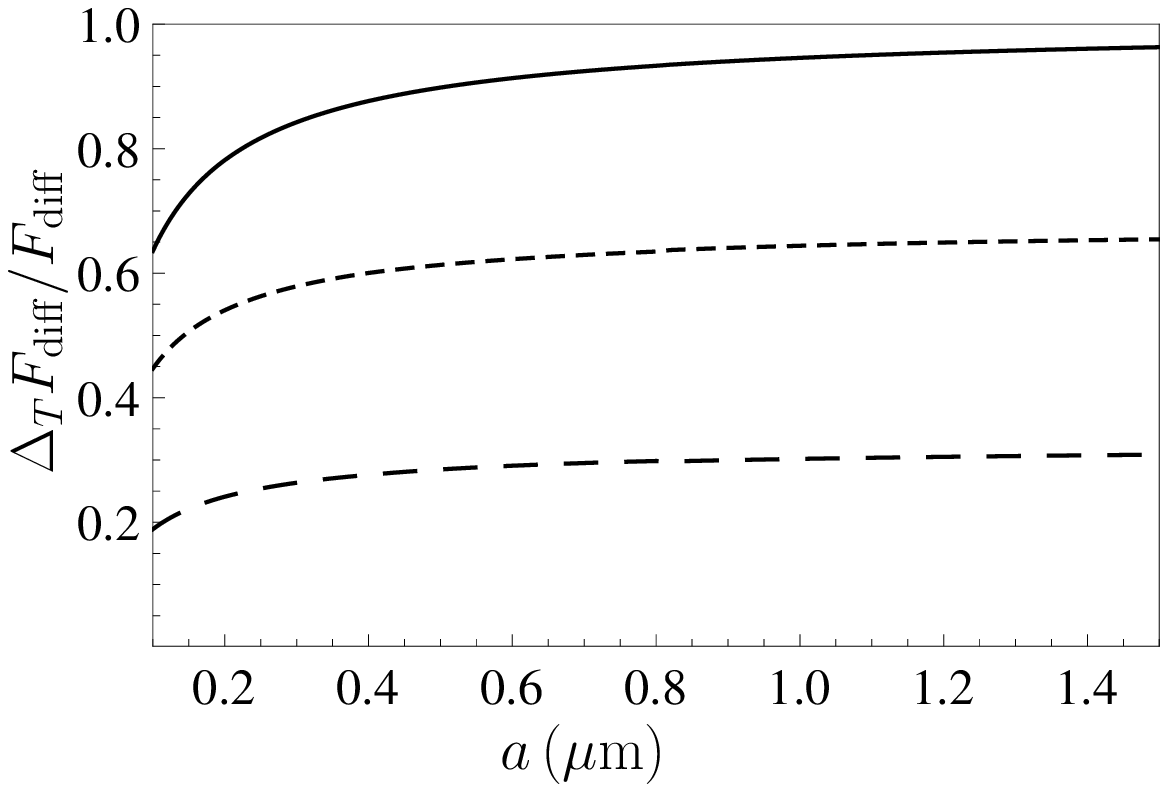}
}
\vspace*{-9cm}
\caption{\label{fg3}
The relative thermal correction $\Delta_TF_{\rm diff}/F_{\rm diff}$ to the difference
among the Casimir forces (solid line).
The short dashed line shows the relative contribution
$\Delta_{T}^{\!(1)}F_{\rm diff}/F_{\rm diff}$ to the thermal correction arising
from the
explicit temperature dependence of the
polarization tensor of graphene, while the long-dashed line  shows the relative contribution $\Delta_{T}^{\!(2)}F_{\rm diff}/F_{\rm diff}$  originating from the combined effect of the temperature dependence of the Au relaxation frequency  together with  the implicit temperature dependence
brought about by the discrete summation over the Matsubara frequencies.
}
\end{figure}

\begin{thebibliography}{99}
\bibitem{1}
M.~I.~Katsnelson, K.~S.~Novoselov, and A.~K.~Geim, Nature Phys. {\bf 2}, 620
(2006).
\bibitem{2}
M.~I.~Katsnelson,
{\it Graphene: Carbon in Two Dimensions}
(Cambridge University Press, Cambridge, 2012).
\bibitem{3}
A.~H.~Castro Neto, F.~Guinea, N.~M.~R.~Peres, K.~S.~Novoselov,
and A.~K.~Geim,
Rev. Mod. Phys. {\bf 81}, 109 (2009).
\bibitem{4}
S.~Das~Sarma, S.~Adam, E.~H.~Hwang, and E.\ Rosa,
Rev. Mod. Phys. {\bf 83}, 407 (2011).
\bibitem{5}
P.~Vogt, P.~De~Padova, C.~Quaresima, J.~Avila,
E.\ Frantzeskakis, M.\ C.\ Asensio, A.\ Resta,
B.\ Ealet, and G.\ L.\ Lay,
Phys. Rev. Lett. {\bf 108}, 155501 (2012).
\bibitem{6}
M.~E.~D\'{a}vila, L.~Xian, S.~Cahangirov, A.\ Rubio,
and G.\ L.\ Lay,
New J. Phys.  {\bf 16}, 095002 (2014).
\bibitem{7}
F.-f.~Zhu, W.-j.~Chen, Y.~Xu, C.-l.~Gao, D.-d.\ Gun,
C.-h.\ Liu, D.\ Qian, S.-C.\ Zhang, and J.-f.\ Jia,
Nature Mater.  {\bf 14}, 1020 (2015).
\bibitem{8}
V.~A.~Parsegian,
{\it Van der Waals Forces: A Handbook for Biologists,
Chemists, Engineers, and Physicists}
(Cambridge University Press, Cambridge, 2005).
\bibitem{9}
M.~Bordag, G.~L.~Klimchitskaya, U.\ Mohideen, and
V.\ M.\ Mostepanenko, {\it Advances in the Casimir Effect}
(Oxford University Press, Oxford, 2015).
\bibitem {10}
J.~Hermann, R.~A.~Distasio, Jr., and A.\ Tkatchenko,
Chem. Reviews {\bf 117}, 4714 (2017).
\bibitem {11}
G.~L.~Klimchitskaya, U. Mohideen, and V.\ M.\ Mostepanenko,
 Rev. Mod. Phys. {\bf 81}, 1827 (2009).
\bibitem{12}
A.~W.~Rodrigues, F.~Capasso, and S.~G.~Johnson, Nature Photonics
{\bf 5}, 211 (2011).
\bibitem{13}
L.~M.~Woods, D.~A.~R.~Dalvit, A.~Tkatchenko, P.\ Rodrigues-Lopez,
A.\ W.\ Rodrigues, and R.\ Podgornik,
Rev. Mod. Phys. {\bf 88}, 045003 (2016).
\bibitem{14}
R.~S.~Decca, D.~L\'opez, E.~Fischbach, G.~L.~Klimchitskaya,
D.~E.~Krause, and V.~M.~Mostepanenko,
Phys. Rev. D {\bf 75}, 077101 (2007).
\bibitem{15}
R.~S.~Decca, D.~L\'opez, E.~Fischbach, G.~L.~Klimchitskaya,
D.~E.~Krause, and V.~M.~Mostepanenko,
Eur. Phys. J. C {\bf 51}, 963 (2007).
\bibitem{16}
C.-C.~Chang, A.~A.~Banishev, R.~Castillo-Garza,
G.~L.~Klimchitskaya, V.\ M.\ Mostepanenko, and U.\ Mohideen,
Phys. Rev. B {\bf 85}, 165443 (2012).
\bibitem{17}
A.~A.~Banishev,
G.~L.~Klimchitskaya, V.\ M.\ Mostepanenko, and U.\ Mohideen,
Phys. Rev. Lett. {\bf 110}, 137401 (2013).
\bibitem{18}
A.~A.~Banishev,
G.~L.~Klimchitskaya, V.\ M.\ Mostepanenko, and U.\ Mohideen,
Phys. Rev. B {\bf 88}, 155410 (2013).
\bibitem{19}
G.~Bimonte, D.~L\'{o}pez, and R.{\ }S.\ Decca,
Phys. Rev. B {\bf 93}, 184434 (2016).
\bibitem{20}
H.{\ }B. Chan, V.{\ }A. Aksyuk, R.{\ }N. Kleiman, D.{\ }J. Bishop, and F. Capasso,
Science {\bf 291}, 1941 (2001).
\bibitem {21}
W.~Broer, H.~Waalkens, V.\ B.\ Svetovoy, J.\ Knoester,
and G.\ Palasantzas,
Phys. Rev. Applied {\bf 4}, 054016 (2015).
\bibitem {22}
M.~Sedighi, W.~H.~Broer, G.\ Palasantzas, and B.\ J.\ Kooi,
Phys. Rev. B {\bf 88}, 165423 (2013).
\bibitem{23}
J.~Zou, Z.~Marcet, A.{\ }W.{\ }Rodriguez, M.\ T.\ H.\ Reid,
A.\ P.\ McCauley, I.\ I.\ Kravchenko, T.\ Lu, Y.\ Bao,
S.{\ }G.\ Johnson, and H.\ B.\ Chan,
Nature Commun. {\bf 4}, 1845 (2013).
\bibitem{24}
G.~G\'{o}mez-Santos,
Phys. Rev. B {\bf 80}, 245424 (2009).
\bibitem{25}
D.~Drosdoff and L.~M.~Woods,
Phys. Rev. B {\bf 82}, 155459 (2010).
\bibitem{26}
Bo~E.~Sernelius,
{Phys. Rev.} B {\bf 85}, 195427 (2012).
\bibitem{27}
A.~D.~Phan, L.~M.~Woods, D.~Drosdoff,
I.\ V.\ Bondarev, and N.\ A.\ Viet,
Appl. Phys. Lett. {\bf 101}, 113118 (2012).
\bibitem{28}
J.~F.~Dobson, T.~Gould, and G.\ Vignale,
{Phys. Rev. X} {\bf 4}, 021040 (2014).
\bibitem {29}
V.\ B.\ Svetovoy and G.\ Palasantzas,
Phys. Rev. Applied {\bf 2}, 034006 (2014).
\bibitem{30}
N.~Knusnutdinov, R.~Kashapov,
 and L.~M.~Woods,
 Phys. Rev. A {\bf 94}, 012513 (2016).
\bibitem{31}
E.~M.~Chudnovsky and R.~Zarzuela,
Phys. Rev. B {\bf 94}, 085424 (2016).
\bibitem{32}
D.~Drosdoff, I.~V.~Bondarev, A.\ Widom, R.\ Podgornik,
 and L.~M.~Woods,
 Phys. Rev. X {\bf 6}, 011004 (2016).
\bibitem{33}
Bo~E.~Sernelius,
{J. Phys.: Condens. Matter} {\bf 27}, 214017 (2015).
\bibitem{34}
P.~Rodriguez-Lopez, W.~J.~M.~Kort-Kamp, D.\ A.\ R.\ Dalvit,
 and L.~M.~Woods,
Nature Commun. {\bf 8}, 14699 (2017).
\bibitem{35}
M.~Bordag, I.~V.~Fialkovsky, D.~M.~Gitman, and
D.~V.~Vassilevich,
{Phys. Rev. B} {\bf 80}, 245406 (2009).
\bibitem{36}
I.~V.~Fialkovsky, V.~N.~Marachevsky, and
D.~V.~Vassilevich,
{Phys. Rev. B} {\bf 84}, 035446 (2011).
\bibitem{37}
M.~Bordag, G.~L.~Klimchitskaya, and
V.\ M.\ Mostepanenko,
Phys. Rev. B {\bf 86}, 165429 (2012).
\bibitem{38}
M.~Chaichian, G.~L.~Klimchitskaya, V.\ M.\ Mostepanenko,
and A.~Tureanu,
Phys. Rev. A {\bf 86}, 012515 (2012).
\bibitem{39}
G.~L.~Klimchitskaya
and V.~M.~Mostepanenko,
{Phys. Rev.} B {\bf 87}, 075439 (2013).
\bibitem{40}
G.~L.~Klimchitskaya, U.~Mohideen, and V.~M.~Mostepanenko,
Phys. Rev. B {\bf 89}, 115419 (2014).
\bibitem{41}
M.~Bordag, G.~L.~Klimchitskaya, V.~M.~Mostepanenko, and V.~M.~Petrov,
Phys. Rev. D {\bf 91}, 045037 (2015); {\bf 93}, 089907(E) (2016).
\bibitem{42}
G.~L.~Klimchitskaya, V.~M.~Mostepanenko, and
Bo~E.~Sernelius,
Phys. Rev. B {\bf 89}, 125407 (2014).
\bibitem{43}
G.~L.~Klimchitskaya
and V.~M.~Mostepanenko,
{Phys. Rev.} A {\bf 89}, 052512 (2014).
\bibitem{44}
G.~L.~Klimchitskaya
and V.~M.~Mostepanenko,
{Phys. Rev.} A {\bf 91}, 174501 (2015).
\bibitem{45}
G. Bimonte, Phys. Rev. Lett. {\bf 112}, 240401 (2014).
\bibitem{46}
G. Bimonte, Phys. Rev. Lett. {\bf 113}, 240405 (2014).
\bibitem{47}
G. Bimonte, Phys. Rev. B {\bf 91}, 205443 (2015).
\bibitem{48}
R.~S.~Decca, D.~L\'opez, E.~Fischbach,
 D.~E.~Krause, and C.~R.~Jamell,
Phys. Rev. Lett. {\bf 94}, 240401 (2005).
\bibitem{49}
Y.-J.~Chen, W.~K.~Tham,
 D.~E.~Krause, D.~L\'opez, E.~Fischbach, and R.~S.~Decca,
{Phys. Rev. Lett.} {\bf 116}, 221102  (2016).
\bibitem{51}
E.~M.~Lifshitz,
Zh. Eksp. Teor. Fiz. {\bf 29}, 94 (1955)
[Sov. Phys. JETP  {\bf 2}, 73 (1956)].
\bibitem{52}
J.~Blocki, J.~Randrup, W.~J.~Swiatecki and C.~F.~Tsang,
{ Ann. Phys. (N.Y.)} {\bf 105}, 427 (1977).
\bibitem{52a}
C.~D.~Fosco, F.~C.~Lombardo, and F.~D.~Mazzitelli,
Phys. Rev. D {\bf 84}, 105031 (2011).
\bibitem{53}
G.~Bimonte, T.~Emig, R.~L.~Jaffe, and M.~Kardar,
Europhys. Lett. {\bf 97}, 50001 (2012).
\bibitem{54}
G.~Bimonte, T.~Emig, and M.~Kardar,
Appl. Phys. Lett. {\bf 100}, 074110 (2012).
\bibitem{55}
L.~P.~Teo,
Phys. Rev. D {\bf 88}, 045019 (2013).
\bibitem{56}
L.~A.~Falkovsky and S.~S.~Pershoguba,
Phys. Rev. B {\bf 76}, 153410 (2007).
\bibitem{57}
T.~Stauber, N.~M.~R.~Peres, and A.~K.~Geim,
Phys. Rev. B {\bf 78}, 085432 (2008).
\bibitem{58}
G.~L.~Klimchitskaya
and V.~M.~Mostepanenko,
{Phys. Rev.} B {\bf 93}, 245419 (2016).
\bibitem{59}
G.~L.~Klimchitskaya
and V.~M.~Mostepanenko,
{Phys. Rev.} B {\bf 94}, 195405 (2016).
\bibitem {60}
{\it Handbook of Optical Constants of Solids},
ed. E.~D.~Palik (Academic, New York, 1985).
\bibitem {61}
{\it Handbook of Optical Constants of Solids}, vol.~2,
ed. E.~D.~Palik (Academic, New York, 1991).
\bibitem{62}
Wang-Kong Tse and S.~Das~Sarma,
Phys. Rev. B {\bf 79}, 235406 (2009).
 \end{thebibliography}
\end{document}